\documentstyle[aps,prb,epsf,multicol]{revtex}
\tighten
\begin{document} 
\title{\Large \bf{
   Dissipation and noise in adiabatic quantum pumps 
   } } 
\author{M. Moskalets$^{1,2}$ and 
M. B\"uttiker$^1$
}
\address{
   $^1$D\'epartement de Physique Th\'eorique, Universit\'e de Gen\`eve,
   CH-1211 Gen\`eve 4, Switzerland\\
    $^2$Department of Metal and Semiconductor Physics,
        National Technic University "Kharkov Polytechnic Institute",
        Kharkov, Ukraine\\}
\date\today
\maketitle
\bigskip

\begin{abstract} 
We investigate the distribution function, the heat flow and the noise
properties of an adiabatic quantum pump for an arbitrary relation 
of pump frequency $\omega$ and temperature. 
To achieve this we start with the scattering matrix 
approach for ac-transport. This approach leads to expressions 
for the quantities of interest in terms of the side bands 
of particles exiting the pump. The side bands correspond 
to particles which have gained or lost a modulation quantum 
$\hbar \omega$. 
We find that our results for the pump current, the heat flow and 
the noise can all be 
expressed in terms of a parametric emissivity matrix. 
In particular we find that the current cross-correlations of a   
multiterminal pump are directly related a to a non-diagonal element
of the parametric emissivity matrix. 
The approach allows 
a description of the quantum statistical correlation 
properties (noise) of an adiabatic quantum pump.

\end{abstract} 
\ \\ 
PACS:  72.10.-d, 73.23.-b \\

\begin{multicols}{2}
\narrowtext

\section{Introduction}
\indent

A resent experiment by Switkes et al. \cite{SMCG99} has stimulated increasing
interest in adiabatic quantum charge pumping. 
Idealy in such an experiment one aims at generating a dc-current by slowly
modulating the shape of a mesoscopic conductor with the help of 
oscillating gate voltages.
A single potential oscillating at frequency $\omega$ does not generate a dc-current,
but two potentials oscillating with the same frequency but out of phase can generate
a dc-current.
The effect is of interest under conditions in which electron motion
is phase-coherent and is thus termed {\it quantum} pumping.
The frequency of the potential modulation is small compared to the characteristic 
times for traversal and reflection of electrons and the pump is thus 
termed {\it adiabatic}.
Thus carriers traversing the sample see an almost static potential.
The last circumstance allows to give
an elegant formulation of quantum pumping \cite{Brouwer98}
which is based on the scattering matrix approach to low frequency ac transport 
in phase coherent mesoscopic systems \cite{BTP94}. \\

Recently Avron et al. \cite{AEGS01} investigated 
adiabatic quantum pumping with the aim to formulate 
criteria for an "optimal pump". 
The term "optimal" means that such a pump is noiseless
and transports integer charge in each cycle.
To this extent they have 
investigated not only the dc-current but also 
the dissipation and the noise generated by a pump. 
Avron et al. express their results in terms 
of an energy shift matrix $i\hbar\partial{\hat s}/\partial t {\hat s}^{\dagger}$
where ${\hat s}$ is the time-dependend scattering matrix. 
This is an elegant formulation
which gives a correct description of time-dependent 
adiabatic currents and dissipation. 
However, for quantities which invoke 
correlations at different times the approach is valid only for pump
frequencies $\hbar \omega << k_BT$. 

It is the purpose of this work to investigate 
the distribution function, heat flow and noise
properties of an adiabatic pump for an arbitrary relation 
of pump frequency and temperature. 
To achieve this we start with the scattering matrix 
approach for ac-transport. This approach leads to expressions 
for the quantities of interest in terms of the side bands 
of particles exiting the pump. The side bands correspond 
to particles which have gained or lost a modulation quantum 
$\hbar \omega$. 
In particular, the approach presented here allows 
a description of the quantum statistical correlation 
properties (noise) of an adiabatic quantum pump.

The adiabatic quantum pump 
\cite{SMCG99,Brouwer98,ZSA99,AK00,SAA00,Simon,WWG00,AAK00,LEWW00a,AEGS00,Brouwer01,PB01,MB01,Levitov01,MM01,SC01,WWGR01,CB01,MCM01,EWAL02}
of interest here should be distinguished from
a variety of other pumping mechanisms.
For certains pumps \cite{Thouless83,Niu90} the charge transferred in 
each cycle is quantized. 
Quantized charge pumping is most easily achieved in devices based on
the Coulomb blockade effect 
\cite{KJVHF91,PLUED92,OKKVH97,AA98,AM01,HWSN01}
where the charge on a quantum dot is quantized.
This is of considerable metrological interest \cite{KMK98,CKKM00}.
Other effects which lead to pumping are
the photovoltaic effect \cite{SZM95,VAA01} and
the acoustoelectric effect 
\cite{SMTGSPR96,TSPSFLRJ97,LEWW00,EWLW01,AEW01,RTPCLR02}.

The paper is organized as follows. 
In Sec.II the essential assumptions we make are described. 
In Sec.III we calculate the nonequilibrium distribution
function for the outgoing particles produced by the pump. 
In Sec.IV we formulate the condition which is necessary to pump dc-current. 
In Sec.V we calculate the heat flows produced 
by an oscillating mesoscopic scatterer. 
In Sec.VI we consider the shot noise produced by the pump and 
analyze the noise in terms of uncorrelated movement  
of nonequilibrium quasi-particles
(quasi-electrons and holes) generated by the pump 
and correlations between them \cite{JB97,BB00}.
In Sec.VII we present explicit results for the particular case of 
a two-leads scatterer 
with the time-reversal symmetry.

\section{The main assumptions} 
\indent

To describe the response of a mesoscopic phase coherent sample 
to slowly oscillating (with a frequency $\omega$)
external real parameters $X_j(t)$ 
(gate potential, magnetic flux, etc.)

\begin{equation} 
  X_j(t) = X_j + X_{\omega,j} e^{i(\omega t - \varphi_j)} 
               + X_{\omega,j} e^{-i(\omega t - \varphi_j)},
\label{Eq1}
\end{equation}

\noindent 
we will use the scattering matrix approach \cite{Buttiker90,Buttiker92,BTP94}.  
The sample is connected via leads 
(which we will number via Greek letters $\alpha$, $\beta$, $\gamma$, etc.)  
to $N_r$ reservoirs.
The scattering matrix $\hat s$ being a function of parameters $X_j(t)$
depends on time.  
Two main assumption will be used.  
First, we suppose that the external parameter changes so slowly 
that we can apply
an "instant scattering" description using the scattering matrix
$\hat s(t)$ frozen at some time $t$.  
Physically this means that the
scattering matrix changes only a little while an electron is scattered by
the mesoscopic sample 
(i.e., the frequency $\omega$ is much smaller
than the inverse Wigner time delay \cite{Wigner55,Smith60}).
In this sense we use the term "adiabatic" pump.

Second, we assume that the amplitude $X_{\omega,j}$ is small
enough to keep only the terms linear in $X_{\omega,j}$ in an
expansion of the scattering matrix

\begin{equation}
 \hat s(t) \approx \hat s + \hat s_{-\omega} e^{i\omega t} 
                          + \hat s_{+\omega} e^{-i\omega t}.
\label{Eq2} 
\end{equation}

\noindent In the limit of small frequencies 
the amplitudes $\hat s_{\pm \omega}$ can be expressed
in terms of parametric derivatives of the on-shell
scattering matrix $\hat s$, 

\begin{equation}
\hat s_{\pm\omega} = \sum_j X_{\omega,j} e^{\pm i\varphi_j} 
\partial\hat s/\partial X_j.
\label{Eq3}
\end{equation}

\noindent
The expansion Eq.(\ref{Eq2}) is equivalent to the nearest sidebands 
approximation \cite{BL82,BTP94} which implies that a scattered electron 
can absorb or emit only one energy quantum $\hbar\omega$ 
before it leaves the scattering region. 

The kinetic properties (charge current, heat current, etc.) which are
of interest here depend on the values of the scattering matrix 
within the energy interval of the order of 
$\max(k_BT,\hbar\omega)$ near the Fermi energy.
In the low frequency ($\omega\to 0$) and low temperature ($T\to 0$) limit
we assume the scattering matrix to be energy independent.

\section{Outgoing distribution function}
\indent

In a pump setup the mesoscopic scatterer is coupled to reservoirs
$\alpha = 1,2,...N_r$
with the same temperatures 
$T_\alpha$ = $T$
and electrochemical potentials
$\mu_\alpha$ = $\mu$. Thus electrons with the energy $E$ 
entering the scatterer are described by the Fermi distribution function
$$
f_\alpha^{(in)}(E) = f_0(E) = \frac{1}{1 + e^{\frac{E-\mu}{k_BT}}}. 
$$
\noindent
Due to the interaction
with an oscillating scatterer an electron can absorb or emit an energy quantum
$\hbar\omega$ that changes the distribution function. 
Our aim is to find the distribution function for outgoing particles 
(i.e., for electrons leaving the mesoscopic sample and entering the reservoir) 
far from the scatterer.

Let us consider a single transverse channel in one of the leads. We introduce
two kinds of carriers \cite{Buttiker92}. First, incoming particles
which are going from the reservoir to the scatterer. And, second, outgoing
particles which are leaving the scattering region. 
We can express the operators $\hat b_\alpha$ which annihilate
outgoing carriers in the lead $\alpha$ 
in terms of operators $\hat a_\beta$ annihilating incoming electrons 
\cite{Buttiker92} in lead $\beta$.
Applying the hypothesis of an instant scattering 
we can write

\begin{equation}
  \hat b_\alpha(t) = \sum_{\beta} s_{\alpha\beta}(t) \hat a_\beta(t).
\label{Eq4}
\end{equation}

\noindent
Here $s_{\alpha\beta}$ is an element of the scattering matrix $\hat s$;
the time dependent operator is 
$\hat a_\alpha (t)$ = $\int dE \hat a_\alpha (E) e^{-iEt/\hbar}$, 
and the energy dependent operators obey the following anticommutation
relations \cite{Buttiker92} 
$$
[\hat a^\dagger_\alpha (E), \hat a_\beta (E')] = 
\delta_{\alpha\beta}\delta(E-E').
$$
\noindent
Note that above expressions correspond to single (transverse) channel leads
and spinless electrons. For the case of many-channel leads 
each lead index ($\alpha$, $\beta$, etc.) includes a transverse channel index
and any repeating lead  index implies implicitly a summation over all 
the transverse channels in the lead. 
Similarly an electron spin can be taken into account.

Using Eq.(\ref{Eq2}) and Eq.(\ref{Eq4}) we obtain \cite{BTP94}

\begin{eqnarray}
 \hat b_\alpha (E) = \sum_\beta 
 s_{\alpha\beta} \hat a_\beta (E)~~~~~~~~~~~~~~~~  \nonumber \\
 + s_{-\omega,\alpha\beta} \hat a_\beta (E+\hbar\omega) +
 s_{+\omega,\alpha\beta} \hat a_\beta (E-\hbar\omega).
\label{Eq5}
\end{eqnarray}

The distribution function for electrons leaving the scatterer through the lead
$\alpha$ is 
$f_{\alpha}^{(out)}(E)$ = $<\hat b_\alpha^\dagger (E)\hat b_\alpha (E)>$,
where $<...>$ means quantum-mechanical averaging.
Substituting Eq.(\ref{Eq5}) we find

\begin{eqnarray}
 f_\alpha^{(out)}(E) = \sum_\beta |s_{\alpha\beta}|^2 f_0(E)~~~~~~~~~~~~ 
 \nonumber \\
 + |s_{-\omega,\alpha\beta}|^2 f_0(E+\hbar\omega) +
 |s_{+\omega,\alpha\beta}|^2 f_0(E-\hbar\omega).
\label{Eq6}
\end{eqnarray}

\noindent Note that the distribution function for outgoing carriers is
a nonequilibrium distribution function 
generated by the nonstationary scatterer. 
The above expression gives a simple physical interpretation for the Fourier
amplitudes of the scattering matrix.
$|s_{-\omega,\alpha\beta}|^2$ 
($|s_{+\omega,\alpha\beta}|^2$) 
is the probability for an electron entering the scatterer through the lead $\beta$
and leaving the scatterer through the lead $\alpha$ to emit (to absorb) 
an energy quantum \cite{BL82} $\hbar\omega$. 
Note that $|s_{\alpha\beta}|^2$ 
is the probability for the same scattering without the change of an energy.
Below we will use Eq.(\ref{Eq6}) to analyze the kinetics of a pump.

\section{dc-current}
\indent

To be definite we take currents from the scatterer to the reservoirs to be positive.
Using the distribution functions $f_0(E)$ for incoming electrons
and $f^{(out)}_\alpha(E)$ for outgoing electrons we find
for the dc-current $I_\alpha$ in the lead 
$\alpha$ far from the scatterer

\begin{equation}
  I_\alpha = \frac{e}{h} \int_0^\infty dE~[f^{(out)}_\alpha(E)-f_0(E)].
\label{Eq7}
\end{equation}

\noindent
Substituting Eq.(\ref{Eq6}) we get

\begin{equation}
  I_\alpha = \frac{e\omega}{2\pi} [T_{+\omega,\alpha} - T_{-\omega,\alpha}].
\label{Eq8}
\end{equation}

\noindent Here we have introduced  
the total probabilities for electrons scattered into the lead $\alpha$ 
(irrespective of the lead through which they entered the scattering region)
to absorb $T_{+\omega,\alpha}$ or to emit $T_{-\omega,\alpha}$ 
an energy quantum $\hbar\omega$

\begin{equation}
T_{\pm\omega,\alpha} = \sum_\beta |s_{\pm\omega,\alpha\beta}|^2.
\label{Eq9}
\end{equation}

\noindent 
We see that only a scatterer with the property

\begin{equation}
 T_{+\omega,\alpha}\neq T_{-\omega,\alpha},
\label{Eq9a}
\end{equation}

\noindent
can pump current into the lead $\alpha$.

It is useful to express 
these probabilities in terms of a bare scattering 
matrix $\hat s$.
To this end we introduce a generalized parametric 
emissivity matrix $\hat\nu[X]$

\begin{equation}
 \hat\nu[X] = -\frac{1}{2\pi i}
  \frac{d\hat s}{dX}\hat s^\dagger.
\label{Eq9b}
\end{equation}

\noindent
with matrix elements

\begin{equation}
 \nu_{\alpha\beta}[X] =  -\frac{1}{2\pi i}
  \sum_\gamma \frac{ds_{\alpha\gamma}}{dX} s^*_{\beta\gamma}.
\label{Eq9ba}
\end{equation}

\noindent 
The diagonal element 
$\nu_{\alpha\alpha}[X]$
of the parametric emissivity 
matrix \cite{Brouwer98,BTP94,Buttiker96,Buttiker01} is the 
charge that leaves the sample through contact $\alpha$ in 
response to a variation of the parameter $X$.
The non-diagonal element 
$\nu_{\alpha\beta}[X]$ ($\alpha \neq \beta$)
of the parametric emissivity matrix 
determines the correlations between current amplitudes generated in 
the contacts $\alpha$ and $\beta$ due to a variation of the parameter $X$
(see Eq.(\ref{Eq24a})).

Using Eq.(\ref{Eq3}) and Eq.(\ref{Eq11}) we express 
the probabilities $T_{\pm\omega,\alpha}$ given by Eq.(\ref{Eq9}) 
in terms of the matrix elements of the parametric emissivity matrix 
(to lowest order in $X_{\omega,j}$)

\begin{equation}
 T_{\pm\omega,\alpha} = 4\pi^2 \sum_\beta \left|
  \sum_j X_{\omega,j} e^{\pm i\varphi_j} \nu_{\alpha\beta}[X_j]
 \right|^2.
\label{Eq9c}
\end{equation}

\begin{figure}
  \vspace{3mm}
  \centerline{
   \epsfxsize4cm
   \epsffile{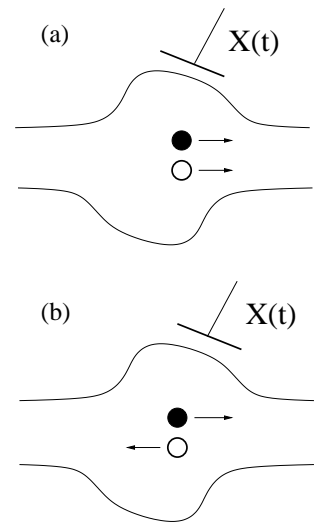}
  }
  \vspace{3mm}
  \nopagebreak
  \caption{
When the parameter $X$ changes, 
the electron system gains energy from the scatterer.
Absorption of an energy quantum $\hbar \omega$ leads to creation 
of nonequilibrium (quasi-)electron-hole pairs.
The electron (black circle) and  hole (open circle) belonging to 
the same pair can be scattered either into one lead (a) or 
into different leads (b).
In the case (a) the quasi-particles do not contribute to the dc current,
but in the case (b) they do contribute.
The process shown in the panel (b) contributes to the current cross-correlations.
In both cases (a) and (b) the quasi-particles carry
energy from the scatterer to the reservoirs.
  }
\label{fig1}
\end{figure}

The quantities $T_{\pm\omega,\alpha}$ admit a simple interpretation 
in the quasi-particle picture. 
Due to scattering the electron system gains an energy from the
nonstationary (oscillating) scatterer. 
Absorption of an energy quantum $\hbar\omega$ leads to creation of
a nonequilibrium (quasi-)electron-hole pair. 
Note that at any temperature $T\neq 0$ 
equilibrium electron-hole pairs exist.
This nonequilibrium pair is neutral but transfers an energy $\hbar\omega$.
From Eq.(\ref{Eq6}) it follows that $T_{+(-)\omega,\alpha}$
is proportional to the number of nonequilibrium quasi-electrons 
(holes) leaving the scattering region through the lead $\alpha$. 
The electron and hole (belonging to the same pair) can be scattered 
either into one lead 
(see Fig.\ref{fig1}.a) 
or into different leads 
(see Fig.\ref{fig1}.b). 
If they are scattered into the same lead they do not contribute to
the current. But if they are scattered into different leads they do contribute.
In any case they contribute to the heat transfer from the oscillating scatterer
into the reservoirs.

The dc-current $I_\alpha$ in the lead $\alpha$ can be represented as 
a sum of two contributions
$I_\alpha$ = $I_\alpha^{(e)}$ + $I_\alpha^{(h)}$,
where 
$I_\alpha^{(e)}$ = $e\omega T_{+\omega,\alpha}/(2\pi)$
and 
$I_\alpha^{(h)}$ = $-e\omega T_{-\omega,\alpha}/(2\pi)$
are currents carried by nonequilibrium quasi-electrons and holes, respectively
(here e (-e) is an electron (a hole) charge).

Now we will show that current is conserved, i.e., 
$\sum_\alpha I_\alpha = 0$. 
To this end we use the fact that the scattering matrix is unitary

\begin{equation}
\hat s(t)\hat s^\dagger(t) = 1.
\label{Eq10}
\end{equation}

For the expansion Eq.(\ref{Eq2}) this leads to the relations

\begin{equation}
\sum_\gamma [s_{\alpha\gamma} s_{\beta\gamma}^* +
s_{-\omega,\alpha\gamma} s_{-\omega,\beta\gamma}^* +
s_{+\omega,\alpha\gamma} s_{+\omega,\beta\gamma}^*] = \delta_{\alpha\beta},
\label{Eq11}
\end{equation}

\begin{equation}
\sum_\gamma s_{\alpha\gamma} s_{-\omega,\beta\gamma}^* =  
- \sum_\gamma s_{\beta\gamma}^* s_{+\omega,\alpha\gamma},
\label{Eq12}
\end{equation}

\begin{equation}
\sum_\gamma s_{\beta\gamma}^* s_{-\omega,\alpha\gamma} =
- \sum_\gamma s_{\alpha\gamma} s_{+\omega,\beta\gamma}^*.
\label{Eq13}
\end{equation}

\noindent Multiplying Eq.(\ref{Eq12}) and Eq.(\ref{Eq13}) by parts,
summing the result over $\alpha$ and taking into account Eq.(\ref{Eq9}) 
we obtain (neglecting the higher powers of 
$s_{\pm\omega,\alpha\gamma}$)

\begin{equation}
 \sum_\alpha T_{-\omega,\alpha} = 
 \sum_\alpha T_{+\omega,\alpha}.
\label{Eq14}
\end{equation}

Using Eqs.(\ref{Eq8}) and (\ref{Eq14}) we see that 
the scatterer does not produce any current
$\sum_\alpha I_\alpha$ = 0 but it can only push 
a current from some reservoir to another reservoir.

An alternative (but equivalent) way to find the dc-current is to average
the time-dependent current 
$$
I_\alpha = \lim\limits_{\Delta t\to\infty} \frac{1}{\Delta t}
\int\limits_0^{\Delta t} dt <\hat I_\alpha(t)>. 
$$
\noindent
The current operator is \cite{Buttiker92}

\begin{equation}
 \hat I_\alpha(t) = \frac{e}{h}[\hat b^\dagger_\alpha(t)\hat b_\alpha(t) -
 \hat a^\dagger_\alpha(t)\hat a_\alpha(t)].
\label{Eq15}
\end{equation}

\noindent
Substituting Eqs.(\ref{Eq2}) and (\ref{Eq4}) into Eq.(\ref{Eq15})
and performing quantum mechanical and time averaging 
we obtain Eq.(\ref{Eq8}). 

Note that in a pump setup where the external reservoirs are
at the same macroscopic conditions 
(electrochemical potential, temperature, etc.)
and a periodic in time perturbation is applied directly to the mesoscopic
conductor there is no linear regime for dc transport
(only ac currents are linear in perturbation).
The dc-currents (charge, heat, etc.) are of a quantum mechanical nature
and arise because of a nonlinear (quadratic) dependence on 
the quantum-mechanical (scattering) amplitudes (see Eq.(\ref{Eq9})).

\section{Heat flow}
\noindent

Particles traversing the sample absorb energy from a time dependent scatterer
and carry it into the reservoirs.
We assume that the reservoirs are
large enough to absorb this energy and to remain
still in thermal equilibrium.
In the leads the energy is transferred by electrons only 
(we neglect any inelastic processes in the leads).
Thus to calculate 
\cite{EA81,SI86,Butcher90,GBJB96,FHK98,Moskalets98,Krive99}
an energy flow $I_{E,\alpha}$ 
entering the reservoir $\alpha$ 
we can use an electron distribution function and write

\begin{equation}
  I_{E,\alpha} = \frac{1}{h} \int_0^\infty dE (E-\mu)
  [f^{(out)}_\alpha(E)-f_0(E)].
\label{Eq16}
\end{equation}

\noindent Substituting Eq.(\ref{Eq6}) we obtain

\begin{equation}
  I_{E,\alpha} = \frac{\hbar\omega^2}{4\pi} 
                [T_{+\omega,\alpha} + T_{-\omega,\alpha}].
\label{Eq17}
\end{equation}

\noindent Comparing Eq.(\ref{Eq8}) and Eq.(\ref{Eq17}) we see that
the time dependent scatterer always generates heat flows
(because $T_{\pm\omega,\alpha}$ are positively defined)
and can be considered as a mesoscopic (phase-coherent) heat source
which can be useful, for instance, for studying various thermoelectric phenomena
in mesoscopic structures.
In contrast the existence of a dc-current Eq.(\ref{Eq8}) requires 
a special condition (see Eq.(\ref{Eq9a})).
Another difference is that the heat flow is directed (at any lead) 
from the scatterer to the reservoir
(if all the reservoirs are at the same temperature)
but the charge flow, if it exists, can be directed 
either from the reservoir to the scatterer (at some lead) or vice versa
(at another lead) because of charge conservation.

The quasi-particle description gives a simple physical interpretation of 
Eq.(\ref{Eq17}). We can say that the heat is transported by two kinds of
quasi-particles, the quasi-electrons and holes. 
Each quasi-particle has an energy $\hbar\omega/2$ (on average). 
This is because the absorption of each energy quantum $\hbar\omega$ creates two 
quasi-particles, a quasi-electron and a hole.
Thus the heat (energy) transferred by quasi-electrons and holes is
$I_{E,\alpha}^{(e)}$ = $(\hbar\omega/2)(I_\alpha^{(e)}/e)$
and
$I_{E,\alpha}^{(h)}$ = $(\hbar\omega/2)(I_\alpha^{(h)}/(-e))$, respectively 
(the quasi-electron $I_\alpha^{(e)}$ and hole $I_\alpha^{(h)}$ currents
are defined in the previous section after Eq.(\ref{Eq9c})). 
The sum of these contributions gives Eq.(\ref{Eq17})

\section{Current fluctuations}
\indent

The problem of current noise in a quantum pump 
is closely connected with the problem of  quantization of the charge 
pumped in one cycle 
\cite{AEGS01,AK00,Levitov01,MM01,AM01}.  
On the other hand the noise in mesoscopic phase coherent conductors 
is interesting in itself 
\cite{BB00,Buttiker90,Buttiker92,Buttiker92a} 
because it is very sensitive to quantum-mechanical interference effects
and can give additional information about the scattering matrix.

To describe the current-current fluctuations we will use 
the correlation function \cite{BB00}

\begin{equation}
 S_{\alpha\beta}(t,t') = \frac{1}{2} 
 <\Delta\hat I_\alpha(t)\Delta\hat I_\beta(t') + 
  \Delta\hat I_\beta(t')\Delta\hat I_\alpha(t)>,
\label{Eq21}
\end{equation}

\noindent where 
$\Delta\hat I$ = $\hat I$ - $<\hat I>$ 
and 
$\hat I_\alpha(t)$
is the quantum-mechanical current operator in the lead $\alpha$ given by Eq.(\ref{Eq15}). 
Note that in the case of a time-dependent scatterer the correlation
function depends on two times $t$ and $t'$.

Here we are interested in the noise averaged over a long time
\cite{LL94,PB98}
($\Delta t\gg 2\pi/\omega$)
and we investigate 
$$
S_{\alpha\beta}(t) = \frac{\omega}{2\pi}
\int\limits_0^{2\pi/\omega} dt' S_{\alpha\beta}(t,t'). 
$$
\noindent
In addition we restrict our consideration to the zero-frequency 
component of the noise spectra
$S_{\alpha\beta}$ = $\int dt S_{\alpha\beta}(t)$.
Substituting the current operator Eq.(\ref{Eq15}) and taking into
account Eq.(\ref{Eq2}) and Eq.(\ref{Eq4}) we can write the zero-frequency
noise power

\begin{eqnarray}
 S_{\alpha\beta} = \frac{2e^2}{h} \int_0^\infty dE
 <\hat S_{\alpha\beta}(E,E)~~~~~~~ \nonumber  \\ 
 + \hat S_{\alpha\beta}(E,E-\hbar\omega) +
 \hat S_{\alpha\beta}(E,E+\hbar\omega)>.
\label{Eq22}
\end{eqnarray}

\noindent Here
$$
\hat S_{\alpha\beta}(E,E') = 
\frac{1}{2}[\Delta\hat I_\alpha(E)\Delta\hat I_\beta(E') +
\Delta\hat I_\beta(E')\Delta\hat I_\alpha(E)];
$$
\noindent
$\Delta\hat I_\alpha(E)$ = $\hat I_\alpha(E)$ - $<\hat I_\alpha(E)>$ and
$$
\hat I_\alpha(E) = \hat b_\alpha^\dagger(E)\hat b_\alpha(E) - 
\hat a_\alpha^\dagger(E)\hat a_\alpha(E).
$$

For the energy independent sattering matrix in the lowest order
in $\hat s_{\pm\omega}$ we obtain
$S_{\alpha\beta}$ = 
$S_{\alpha\beta}^{(th)}$ + $S_{\alpha\beta}^{(pump)}$.
Here the thermal (or Nyquist-Johnson) noise is 
\cite{Buttiker90,BB00}
$$
S_{\alpha\beta}^{(th)} =
2e^2k_BT/h[2\delta_{\alpha\beta} - |s_{\alpha\beta}|^2 - 
|s_{\beta\alpha}|^2 ].
$$
\noindent
The  noise power produced by the pump is

\begin{equation}
 S_{\alpha\beta}^{(pump)} = \frac{2e^2}{h} F(\hbar\omega,k_BT)
 \left( 
 \delta_{\alpha\beta}[T_{-\omega,\alpha} + T_{+\omega,\alpha}] - 
 T_{\alpha\beta}^{(cor)} 
 \right),
\label{Eq23}
\end{equation}

\noindent where

\begin{equation}
 T_{\alpha\beta}^{(cor)} = 
  \left|\sum_\gamma s_{\beta\gamma}s^*_{-\omega,\alpha\gamma}\right|^2 +
  \left|\sum_\gamma s_{\beta\gamma}s^*_{+\omega,\alpha\gamma}\right|^2,
\label{Eq24}
\end{equation}

\noindent and
$F(\hbar\omega,k_BT)$ = $\hbar\omega\coth[\hbar\omega/(2k_BT)] - 2k_BT$.

In addition to the probabilities 
$T_{\pm\omega,\alpha}$ 
which determine the dc current Eq.(\ref{Eq8}) and heat flow Eq.(\ref{Eq17})
there appears a third  key quantity 
$T_{\alpha\beta}^{(cor)}$, 
which describes the effect of correlations 
between (quasi-)particles.
Similarly to 
$T_{\pm\omega,\alpha}$ (see Eqs.(\ref{Eq9c}))
this probability  can be expressed in terms of a generalized emissivity matrix 
$\hat\nu$ (see Eq.(\ref{Eq9b}))

\begin{equation}
 T_{\alpha\beta}^{(cor)} = 4\pi^2 \sum_{\eta = +1,-1}
 \left| \sum_j X_{\omega,j} e^{i\eta\varphi_j} \nu_{\alpha\beta}[X_j]
 \right|^2.
\label{Eq24a}
\end{equation} 
Note that there is no summation over $\alpha$ or $\beta$:
Consequently this probability which determines the current cross-correlation
is directly proportional to the off-diagonal element of the 
emissivity matrix. 

Now we will analyze the noise power Eq.(\ref{Eq23}).
We  can see that the current cross-correlations 
$S_{\alpha\beta}^{(pump)}$ ($\alpha \neq \beta$)
produced by the pump 
are negative: that is quite general for nonequilibrium noise in the system
of fermions \cite{Buttiker92}.
The noise generated by the pump obeys the following sum rules \cite{Buttiker92} 
$\sum_\alpha S_{\alpha\beta}^{(pump)}$ = 
$\sum_\beta S_{\alpha\beta}^{(pump)}$ = 0.
This is a straightforward consequence of an instant scattering description
applied here. 
Indeed, using Eqs.(\ref{Eq4}),(\ref{Eq10}) and (\ref{Eq15}) we can 
see that the conservation law holds not only for a quantum averaged
current $<\hat I_\alpha(t)>$ but for a current operator as well 
\cite{Buttiker92}: $\sum_\alpha \hat I_\alpha(t) = 0$. Thus the 
current correlations 
${\it S_{\alpha\beta}}\sim <\hat I_\alpha\hat I_\beta> $ must obey the same
sum rule.

The function $F(\hbar\omega,k_BT)$ 
describes the effect of thermal fluctuation on shot noise 
and determines the dependence of the noise 
on the pump frequency $\omega$. 
At sufficiently high temperature $\hbar\omega \ll k_BT$ 
the noise Eq.(\ref{Eq23}) is quadratic in $\omega$. 
This is in agreement with 
Ref.\onlinecite{AEGS01}. 
But at low temperature $k_BT \ll \hbar\omega$
the noise is linear in $\omega$ and this is in agreement with 
the counting statistics calculations of Levitov \cite{Levitov01}.

Next consider the three terms in the brackets of the r.h.s. of Eq.(\ref{Eq23}).
Consider the low temperature limit
$k_BT\ll\hbar\omega$ and devide the expression for noise into two parts
$S_{\alpha\beta}^{(pump)}$ = 
$\delta_{\alpha\beta}S_{\alpha}^{(pump),(P)}$ + 
$S_{\alpha\beta}^{(pump),(cor)}$.
The first part
$$
S_{\alpha}^{(pump),(P)} =  
\frac{e^2\omega}{\pi}[T_{-\omega,\alpha} + T_{+\omega,\alpha}],
$$
\noindent
is due to an uncorrelated movement of nonequilibrium quasi-electrons
and holes. To verify this we apply 
the Schottky formula \cite{Schottky18} for shot noise
$S_{\alpha,q}^{(Sch)}$ = $2qI_\alpha^{(q)}$
(here $q$ is a particle charge and the index "q" means that the current
is carried by the particles with the charge q).
Substituting the current carried by the quasi-electrons 
$I_\alpha^{(e)}$ = $e\omega T_{+\omega,\alpha}/(2\pi)$  
and by the holes
$I_\alpha^{(h)}$ = $-e\omega T_{+\omega,\alpha}/(2\pi)$
(here e(-e) is an electron (hole) charge)
into Schottky's formula we obtain 
$S_{\alpha}^{(pump),(P)}$ = $S_{\alpha,e}^{(Sch)}$ + $S_{\alpha,h}^{(Sch)}$
(in the literature the Schottky result is referred as 
the Poisson value of shot noise that we indicate by the upper index "P").

The second part
$$
S_{\alpha\beta}^{(pump),(cor)} = -\frac{e^2\omega}{\pi}T_{\alpha\beta}^{(cor)},
$$
\noindent
is due to correlations between quasi-electrons and holes.
These correlations are a consequence of the common origin of 
the electron and hole forming a pair and their  
subsequent scattering into different leads Fig.\ref{fig1}.b. 
Thus we can say that the cross-correlations are exclusively due to 
dissolving (neutral) electron-hole pairs. 
Because of charge conservation this gives a simple explanation of a negative
sign of cross-correlations in our case Eq.(\ref{Eq23}).
Note that Schottky's result gives no correlations between 
currents at different leads.
Due to 
$S_{\alpha\alpha}^{(pump),(cor)}$
the current correlation at the same lead
$S_{\alpha\alpha}^{(pump)}$ is below the Poisson value
$S_{\alpha}^{(P)}$
and the Fano factor characterizing the deviation of the actual shot noise 
from the Poisson noise (see, e.g., \cite{BB00}) 
$F$ = $S_{\alpha\alpha}^{(pump)}/S_{\alpha}^{(P)}$ is, in general, less than unity.
We would like to emphasize that when we calculate the Poisson value of shot noise
we do not use the total current $I_\alpha$ in the lead $\alpha$
but we calculate the sum of the Poisson noises produced by both 
the quasi-electrons (the current is $I_\alpha^{(e)}$)
and the holes (the current is $I_\alpha^{(h)}$).

\section{Applications}
\noindent

In this section we consider a simple but a quite generic case  
of a two-terminal mesoscopic conductor with a time-reversal symmetry 
(without magnetic fluxes).  
In addition we assume that the external, time-dependent parameters $X_j(t)$
do not change the total charge on a sample 
(if this is not the case we need to take into account 
the self-consistent internal potential \cite{BTP94}). 
In this case we can choose the scattering matrix as follows
(we consider spinless electrons)

\begin{equation}
\hat s = \left(
  \begin{array}{cc}
       re^{i\theta}   & it            \\
       it             & re^{-i\theta} \\
  \end{array}
\right).
\label{Eq25}
\end{equation}

\noindent Here $r^2$ and $t^2$ are the reflection and the transmission 
probability, respectively ($r^2+t^2=1$).
We assume the quantities $r,t,\theta$ to be the functions 
of external parameters $X_j$ (see Eq.(\ref{Eq1})).

\subsection{Heat flow and noise in one-parameter "pumps"}
\indent

If only one external parameter $X$ is varied then
we get from Eq.(\ref{Eq3}) 
$|s_{-\omega,\alpha\beta}|^2$ = $|s_{+\omega,\alpha\beta}|^2$
for any $\alpha$ and $\beta$. Thus the one-parameter adiabatic "pump" 
gives the same probability for absorption and emission at both leads ($\alpha=1,2$)

\begin{eqnarray}
  T_{-\omega,\alpha}^{(1)} = T_{+\omega,\alpha}^{(1)} = 
  T_{\omega,\alpha}^{(1)}, \nonumber \\
  \label{Eq26} \\
  T_{\omega,\alpha}^{(1)}[X] = X_\omega^2\sum_\beta
 \left|\frac{d s_{\alpha\beta}}{d X}\right|^2, \nonumber
\end{eqnarray}

\noindent 
and does not produce a dc-current:
$I_\alpha = 0$ (see Eq.(\ref{Eq8})).
In contrast it does produce the heat flows Eq.(\ref{Eq17}) 
and the noise Eq.(\ref{Eq23}).
The contribution to the noise Eq.(\ref{Eq24})
due to correlations between quasi-particles is

\begin{equation}
  T_{\alpha\beta}^{(cor)} = 8\pi^2X_\omega^2
  \left| \nu_{\alpha\beta}[X]\right|^2.
\label{Eq27}
\end{equation}

\noindent 
Here $\nu_{\alpha,\beta}[X]$ is a matrix element of a generalized 
parametric emissivity matrix $\hat\nu[X]$ Eq.(\ref{Eq9b}).
With the scattering matrix Eq.(\ref{Eq25}) 
we find a heat flow

\begin{eqnarray}
 I_{E,1}^{(1)} = I_{E,2}^{(1)}~~~~~~~~~~~~~~~~~~~~~ \nonumber \\
 =  \frac{\hbar\omega^2 X_\omega^2}{2\pi}
 \left( 
        r^2\left(\frac{d\theta}{dX}\right)^2 +
        \left(\frac{dr}{dX}\right)^2 +
        \left(\frac{dt}{dX}\right)^2
 \right),
\label{Eq29}
\end{eqnarray}

\noindent and a noise 
$S_{11}^{(pump)}$ = $S_{22}^{(pump)}$ = $- S_{12}^{(pump)}$ = 
$- S_{21}^{(pump)}$ = $S^{(1)}$

\begin{eqnarray}
 S^{(1)} = \frac{4e^2 X_\omega^2}{h} F(\hbar\omega,k_BT) 
 ~~~~~~~~~~ \nonumber \\
 \times\left(r^2t^2\left(\frac{d\theta}{dX}\right)^2 + 
 \left(\frac{dr}{dX}\right)^2 +
 \left(\frac{dt}{dX}\right)^2
 \right).
\label{Eq30}
\end{eqnarray}

\noindent 
We see that the noise produced by the one-parameter "pump"
Eq.(\ref{Eq30}) gives us direct information on the dependence 
of the scattering matrix on the varying external parameter $X$. 
This dependence $\hat s(X)$ is important for calculating the current produced
by the pump if two external parameters are varied Eq.(\ref{Eq35}).
In a real experimental situation the dependence $\hat s(X)$ 
is unknown and can not be calculated in a simple way. 
Thus the possibility to obtain this dependence from the experimental data
seems useful.

Note that for some particular conditions the noise and the heat flow 
at the same lead are related by a simple relation.
For instance, if the amplitude $r$ of a reflection coefficient 
is independent of $X$ 
the ratio
$S^{(1)}$/$I^{(1)}_{E}$
at low temperature ($k_BT\ll\hbar\omega$) is 
$8\pi G/\omega$, 
where $G=e^2t^2/h$ is the conductance of our mesoscopic sample.

On the other hand if the phase $\theta$ of a reflection coefficient
is independent of the varying parameter $X$ 
the contribution of quasi-particle correlations 
to the current correlations at the same lead vanishes, i.e.,
$T_{\alpha\alpha}^{(cor)} = 0$.
In this case the noise
$S^{(pump)}_{\alpha\alpha}$
reaches the Poisson value (the Fano factor is $F = 1$) 
and the ratio of the noise to the heat flow is a universal function of 
the temperature and the pump frequency 
$$
S^{(1)}/I^{(1)}_{E} = F^{(1)}(\hbar\omega,k_BT) = 
4e^2 F(\hbar\omega,k_BT)/(\hbar\omega)^2,
$$
\noindent 
which is independent of individual features of a scatterer.
Comparing Eq.(\ref{Eq17}) and Eq.(\ref{Eq23}) we see that this conclusion
is a quite general feature of a weak amplitude pump:
if $T_{\alpha\alpha}^{(cor)} = 0$ the ratio
$S^{(pump)}_{\alpha\alpha}/I_{E,\alpha}$ = $F^{(1)}(\hbar\omega,k_BT)$.
Note that even if the phase $\theta$ is independent of $X$,
$T_{\alpha\beta}^{(cor)} \neq 0$ ($\alpha \neq \beta$) and 
the quasi-particle correlations are important for the current cross-correlations.
To illustrate this fact in the next subsection  we consider  
a particular case of a multiterminal (three-terminal) conductor.
We investigate a case when 
the phase of the transmission (reflection) amplitude is unimportant
but the current cross-correlations are present.

\subsection{Noise and heat flow of an oscillating wave splitter}
\indent

Let us consider a wave splitter in which one lead $\alpha = 1$ 
couples via a tunnel barrier with transparency $\epsilon$
symmetrically to two leads $\alpha = 2, 3$. 
We assume that this three lead structure
is described by the single parameter scattering matrix \cite{BIA84}

\begin{equation}
\hat s = \left(
  \begin{array}{ccc}
       -(a+b) & \sqrt{\epsilon} & \sqrt{\epsilon} \\
       \sqrt{\epsilon}  &  a  &  b  \\
       \sqrt{\epsilon}  &  b  &  a         
   \end{array}
\right).
\label{Eq30a}
\end{equation}

\noindent
where 
$a = (\sqrt{1-2\epsilon} - 1)/2$ and
$b = (\sqrt{1-2\epsilon} + 1)/2$.
For $\epsilon = 0$ carriers incident from lead $1$ are completely reflected, 
for $\epsilon = 1/2$ carriers incident from lead $1$
are transmitted (without reflection) with equal probability into leads $2$
and $3$. 

We choose the transparency $\epsilon$ as an external parameter and 
assume that it is subject to small amplitude oscillations
$\epsilon(t) = \epsilon + 2\epsilon_\omega\cos(\omega t)$
($\epsilon_\omega \ll \epsilon$).
With the scattering matrix Eq.(\ref{Eq30a}) the parametric emissivity matrix
$\hat\nu[\epsilon]$ Eq.(\ref{Eq9b}) is

\begin{equation}
\hat\nu[\epsilon] = \frac{1}{4\pi\sqrt{(1-2\epsilon)\epsilon}}
    \left(
  \begin{array}{ccc}
       0 & i & i  \\
      -i & 0 & 0  \\
      -i & 0 & 0  
   \end{array}
\right).
\label{Eq30b}
\end{equation}

Using Eqs.(\ref{Eq9c}) and (\ref{Eq24a})
we can calculate $T_{\pm\omega,\alpha}$ and $T_{\alpha\beta}^{(cor)}$.
Substituting these probabilities into Eqs.(\ref{Eq8}), (\ref{Eq17}) and 
(\ref{Eq23})
we obtain the quantities of interest here.
We see that the dc charge current $I_\alpha$ is zero at all leads
(as it is expected because only one parameter is varied). 
However the heat flow $I_{E,\alpha}$ and zero-frequency noise power
show interesting features.

Despite the fact that lead $\alpha = 1$ is coupled only weakly, 
the heat flow 
$I_{E,\alpha}$ 
(for any $\epsilon > 0$) and the shot noise power 
$S_{\alpha\alpha}^{(pump)}$
are the same in all three leads

\begin{equation}
 I_{E,1} = I_{E,2} = I_{E,3} = 
 \frac{\hbar\omega^2\epsilon_\omega^2}{4\pi\epsilon(1-2\epsilon)}.
\label{Eq30c}
\end{equation}

\noindent
The heat flow is related to the noise in the simple way discussed at the end 
of the previous subsection
$S_{\alpha\alpha}^{(pump)}/I_{E,\alpha}$ = 
$4e^2F(\hbar\omega,k_BT)/(\hbar\omega)^2$.

The asymmetry between leads appears in cross-correlations
$S^{(pump)}_{\alpha\beta}$ ($\alpha \neq \beta$). 
The cross-correlation of current fluctuations at the two 
symmetrically coupled leads vanishes
$S_{23}^{(pump)} = 0$
but the cross-correlations invoking the fluctuating current
of the weakly coupled lead $1$ are non-vanishing, 
$S_{12}^{(pump)}$ = 
$S_{13}^{(pump)}$ = $- \frac{1}{2}S_{\alpha\alpha}^{(pump)}$.

\subsection{Heat flow and noise in two parameter pumps} 
\indent

Now we return to the scattering matrix Eq.(\ref{Eq25}) and
assume that it depends on
two parameters $X_1(t)$ and $X_2(t)$. 
In this case we can represent the probabilities for absorption and emission
in terms of a symmetric and antisymmetric contribution

\begin{equation}
 T_{\pm\omega,\alpha}^{(2)} = T_{\omega,\alpha}^{(s)} \pm 
  T_{\omega,\alpha}^{(a)}.
\label{Eq31}
\end{equation}

\noindent 
Here the symmetric 
(with respect to absorption and emission of a modulation quantum $\hbar\omega$)
$T_{\omega,\alpha}^{(s)}$ 
and antisymmetric $T_{\omega,\alpha}^{(a)}$ parts 
which determine the heat flow Eq.(\ref{Eq17}) 
and the dc-current Eq.(\ref{Eq8}), respectively, are

\begin{eqnarray}
 T_{\omega,\alpha}^{(s)} = 
 T_{\omega,\alpha}^{(1)}[X_1] + T_{\omega,\alpha}^{(1)}[X_2] 
 ~~~~~~~~~~ \nonumber \\
 + 2X_{\omega,1}X_{\omega,2}\cos(\varphi_2 - \varphi_1)Re[\Pi_\alpha(X_1,X_2)],
\label{Eq32}
\end{eqnarray}

\begin{equation}
 T_{\omega,\alpha}^{(a)} = 
2X_{\omega,1}X_{\omega,2}\sin(\varphi_2 - \varphi_1)
 Im[\Pi_\alpha(X_1,X_2)],
\label{Eq33}
\end{equation}

\noindent with the quantity $\Pi_\alpha(X_1,X_2)$ being

\begin{eqnarray}
 \Pi_\alpha(X_1,X_2) = 
 \sum_\beta \left[ 
\frac{\partial s^*_{\alpha\beta}}{\partial X_1}
\frac{\partial s_{\alpha\beta}}{\partial X_2}\right] 
~~~~~~ \nonumber \\
= 4\pi^2\sum_\beta \nu^*_{\alpha\beta}[X_1] \nu_{\alpha\beta}[X_2].
\label{Eq34}
\end{eqnarray}

\noindent In Eq.(\ref{Eq32}) the quantity $T_{\omega,\alpha}^{(1)}[X]$ is given by
Eq.(\ref{Eq26}).

Substituting Eq.(\ref{Eq31}) into Eq.(\ref{Eq8}) we immediately 
reproduce the result obtained by Brouwer \cite{Brouwer98} for the pumped
current (at small amplitudes $X_{\omega,j}$)

\begin{equation}
 I_\alpha = \frac{2e\omega}{\pi}X_{\omega,1} X_{\omega,2}
  \sin(\varphi_2 - \varphi_1) Im[\Pi_\alpha(X_1,X_2)]. 
\label{Eq35}
\end{equation}

We see that varying only one (no matter which) parameter we can not
generate a dc-current.
But if at least two parameters
$X_1$ and $X_2$ are varied periodically but out of phase
$\varphi_1\neq\varphi_2$ then the mesoscopic sample can continuously
pump charge between reservoirs with the same chemical potentials.
This is a consequence of quantum-mechanical interference effects.
These effects manifest themselves not only in the dc-current 
but also in the heat flow and in the noise. 

To characterize the contribution of interference effects to 
the heat flow we consider the difference 
$\Delta I_{E,\alpha}$
between the heat flow $I_{E}^{(2)}$ when the two parameters
are varied simultaneously and the sum of heat flows when only one parameter
oscillates $I_{E,\alpha}^{(1)}[X_1]$ + $I_{E,\alpha}^{(1)}[X_2]$.
This difference is

\begin{equation}
 \Delta I_{E,\alpha} = 
 \frac{\hbar\omega^2}{\pi} 
X_{\omega,1}X_{\omega,2}\cos(\varphi_2 - \varphi_1)
 Re[\Pi_\alpha(X_1,X_2)].
\label{Eq36}
\end{equation}

\noindent We see that the additional heat production 
$\Delta I_{E,\alpha}$ and the dc-current give a full
description of the quantity $\Pi_\alpha(X_1,X_2)$, i.e., they determine
the real and the imaginary parts.

For the scattering matrix Eq.(\ref{Eq25}) we get

\begin{eqnarray}
  Re[\Pi_1(X_1,X_2)] = Re[\Pi_2(X_1,X_2)]  
  ~~~~~~ \nonumber \\
  = r^2\frac{\partial\theta}{\partial X_1} \frac{\partial\theta}{\partial X_2} +
  \frac{\partial r}{\partial X_1} \frac{\partial r}{\partial X_2} +
  \frac{\partial t}{\partial X_1} \frac{\partial t}{\partial X_2},
\label{Eq37}
\end{eqnarray}

\begin{eqnarray}
  Im[\Pi_1(X_1,X_2)] = -Im[\Pi_2(X_1,X_2)] 
  \nonumber \\
  = \frac{1}{2} \left(
  \frac{\partial r^2}{\partial X_1} \frac{\partial\theta}{\partial X_2} -
  \frac{\partial\theta}{\partial X_1} \frac{\partial r^2}{\partial X_2}
 \right).
\label{Eq38}
\end{eqnarray}

As we did for the heat production we calculate an additional noise
$\Delta S_{\alpha\beta}^{(pump)}$ generated by two simultaneously 
oscillating parameters
$X_1$ and $X_2$ 
over the sum of noises produced by each of them separately,

\begin{eqnarray}
 \Delta S_{\alpha\beta}^{(pump)} = \frac{8e^2}{h} 
   X_{\omega,1} X_{\omega,2} F(\hbar\omega,k_BT)
   \nonumber \\
   \times\cos(\varphi_2 - \varphi_1)
   Re [N_{\alpha\beta}(X_1,X_2)],
\label{Eq39}
\end{eqnarray}

\noindent where

\begin{equation}
N_{\alpha\alpha}(X_1,X_2) = 4\pi^2\sum_{\gamma\neq\alpha} 
   \nu^*_{\alpha\gamma}[X_1] \nu_{\alpha\gamma}[X_2],
\label{Eq40}
\end{equation}

\begin{equation}
N_{\alpha\beta}(X_1,X_2) = - 4\pi^2
      \nu^*_{\alpha\beta}[X_1]\nu_{\alpha\beta}[X_2],
      ~~~~~~\alpha\neq\beta.
\label{Eq41}
\end{equation}

\noindent 
Note the cosine dependence of the additional heat and noise on the phase difference 
$\Delta\varphi$ = $\varphi_2 - \varphi_1$. 
In contrast the pumped current Eq.(\ref{Eq35}) is determined by the sine of the
phase difference $\Delta\varphi$. 
As a consequence if the pumped current is large
(as a function of $\Delta\varphi$) the additional noise and heat flow are small.

For the scattering matrix Eq.(\ref{Eq25}) we have
$Re [N_{11}]$ = $Re [N_{22}]$ = -$Re [N_{12}]$ = -$Re [N_{21}]$ = $Re [N^{(2)}]$

\begin{eqnarray}
Re [N^{(2)}(X_1,X_2)] = 
r^2t^2\frac{\partial\theta}{\partial X_1}\frac{\partial\theta}{\partial X_2} 
\nonumber \\
  + \frac{\partial r}{\partial X_1} \frac{\partial r}{\partial X_2} 
  + \frac{\partial t}{\partial X_1} \frac{\partial t}{\partial X_2}.
\label{Eq42}
\end{eqnarray}

From Eq.(\ref{Eq37}) and Eq.(\ref{Eq42}) we can see that the additional
heat production and the additional noise vanish if 
the amplitude of the reflection coefficient $r$ and its phase $\theta$
depend on a single parameter only (not necessary the same). 
However there remains a heat production Eq.(\ref{Eq29}) and 
noise Eq.(\ref{Eq30}) owing to independently oscillating parameters.
As it is evident from Eq.(\ref{Eq26}) this unavoidable heat production
(see Eq.(\ref{Eq17})) and noise (see Eq.(\ref{Eq23})) are present 
always if only the "pump" is working. 
On the other hand if the phase $\theta$ of the reflection coefficient
depends on only one varying parameter the ratio of 
additional noise Eq.(\ref{Eq39}) to additional
heat production Eq.(\ref{Eq36}) 
does not depend on the scattering matrix 
and is equal to $4e^2 F(\hbar\omega,k_BT)/(\hbar\omega)^2$.

Under some conditions the additional noise 
$\Delta S^{(pump)}_{\alpha\alpha}$
can be related to the dc-current $I_\alpha$ at the same lead.
If $r=r(X_1)$ (or $r=r(X_2)$) then their ratio  
$$
\frac{\Delta S^{(pump)}_{\alpha\alpha}}{I_\alpha} = 
(-)4et^2\cot(\varphi_2-\varphi_1)
\frac{\partial\theta}{\partial r^2}
\frac{F(\hbar\omega,k_BT)}{\hbar\omega},
$$
\noindent
is independent of the varying parameters. 
On the other hand if
$\theta=\theta(X_2)$ (or $\theta=\theta(X_1)$) then we get 
$$
\frac{\Delta S^{(pump)}_{\alpha\alpha}}{I_\alpha} = 
(-)\frac{e}{t^2r^2}\cot(\varphi_2-\varphi_1)
\frac{\partial r^2}{\partial\theta}
\frac{F(\hbar\omega,k_BT)}{\hbar\omega}.
$$
\noindent
In the low temperature limit this ratio becomes independent of frequency.

\section{Conclusion}
\noindent

In this work we have developed the approach to the kinetics 
of an adiabatic quantum pump for an arbitrary relation 
of pump frequency and temperature. 
Our consideration is based on the scattering matrix approach for ac-transport. 
This approach takes into account the existence of
the side bands of particles exiting the pump
and thus allows 
a description of the quantum statistical correlation 
properties (e.g., noise) of an adiabatic quantum pump.  
The side bands correspond to particles which have gained or lost 
a modulation quantum $\hbar \omega$.
We find that our results for the pump current, the heat flow and 
the noise can all be 
expressed in terms of a parametric emissivity matrix. 
In particular we find that the current cross-correlations of a   
multiterminal pump are directly related a to a non-diagonal element
of the parametric emissivity matrix. 

Using the quasi-particle picture
we have given a simple physical interpretations of processes leading to charge
and energy transfer in the system. 
Due to the oscillations of the scatterer 
the electron system gains energy (the side bands arise).  
Absorption of an energy quantum $\hbar\omega$
leads to the creation of a nonequilibrium (quasi-)electron-hole pair.
These quasi-particles carry energy from the scatterer to the reservoirs. 
On average the electron-hole pair is neutral 
thus the pump is not a source of a charge current
but under some conditions \cite{Brouwer98} can only push charge 
from some reservoirs to others. 
These conditions can be realized if
the quasi-electron and hole (belonging to the same pair) 
leave the scatterer through different leads
(say, $\alpha$ and $\beta$).
In this case these quasi-particles contribute to the charge transfer between
the reservoirs $\alpha$ and $\beta$.
These quasiparticles are correlated  
since they are created in the same event. 
This is also the source of the correlations between the currents 
at the leads $\alpha$ and $\beta$ (cross-correlations). 
Thus we conclude that the existence of the dc-current in a weak 
amplitude pump 
is always accompanied by current correlations (shot noise). 
This type of a pump can not be optimal (in particular, noiseless)
in the sense of Ref.\cite{AEGS01}. 

To assess the possibility of an optimal adiabatic quantum pump
further investigations are necessary. 
In particular large amplitude variations of the external parameter
have to be considered. 
We hope that by taking into account many photon processes 
(see, e.g., \cite{PB98})
the approach developed in the present paper can be generalized
to the case of a strong amplitude adiabatic quantum pump.

\ \\
\centerline{\bf ACKNOWLEDGMENTS}
\indent

M. B. thanks G. M. Graf for stimulating discussions.  
Both of us acknowledge  useful discussions with A. Alekseev. 
M.M. gratefully acknowledges the support of the University of Geneva
where this work was done.
This work was supported by the Swiss National Science Foundation.

\end{multicols}
\end{document}